 \documentclass[final,3p,times]{elsarticle}

\usepackage{amssymb}
\usepackage{amsthm}

\usepackage{lineno}
\usepackage{graphicx}
\usepackage{dcolumn}
\usepackage{bm}
\newcommand{\be}{\begin{equation}}
\newcommand{\ee}{\end{equation}}
\newcommand{\bea}{\begin{eqnarray}}
\newcommand{\eea}{\end{eqnarray}}
\newcommand{\ba}{\begin{array}}
\newcommand{\ea}{\end{array}}
\newcommand{\bi}{\begin{itemize}}
\newcommand{\ei}{\end{itemize}}

\newcommand{\nslash}{\kern 0.2 em n\kern -0.50em /}
\newcommand{\kslash}{\kern 0.2 em k\kern -0.45em /}
\newcommand{\qslash}{\kern 0.2 em q\kern -0.45em /}
\newcommand{\pslash}{\kern 0.2 em p\kern -0.50em /}
\newcommand{\rslash}{\kern 0.2 em r\kern -0.50em /}
\newcommand{\sslash}{\kern 0.2 em s\kern -0.50em /}
\newcommand{\Sslash}{\kern 0.2 em S\kern -0.50em /}
\newcommand{\Pslash}{\kern 0.2 em P\kern -0.50em /}
\newcommand{\Dslash}{\kern 0.2 em D\kern -0.65em /\kern 0.15em}

\journal{Physics Letters B}

\begin{document}

\begin{frontmatter}



\title{Compton scattering off proton in the third resonance region}
\author{
Xu Cao$^{a,d}${\footnote{Corresponding author: caoxu@impcas.ac.cn}},
H. Lenske$^{b,c}$
}
\address{
$^a$ Institute of Modern Physics, Chinese Academy of Sciences, Lanzhou 730000, China \\
$^b$ Institut f\"{u}r Theoretische Physik, Universit\"{a}t Giessen, D-35392 Giessen, Germany \\
$^c$ GSI Darmstadt, D-64291 Darmstadt, Germany \\
$^d$ State Key Laboratory of Theoretical Physics, Institute of Theoretical Physics, Chinese
Academy of Sciences, Beijing 100190, China
}

\begin{abstract}
  Compton scattering off the proton in the third resonance region is analyzed for the first time, owing to the full combined analysis of pion- and photo-induced reactions in a coupled-channel effective Lagrangian model with K-matrix approximation. Two isospin $I=3/2$ resonances $D_{33}(1700)$ and $F_{35}(1930)$ are found to be essential in the range of 1.6 - 1.8 GeV. The recent beam asymmetry data of Compton scattering from the GRAAL facility are used to determine the helicity couplings of these resonances, and strong constraints are coming also from $\pi N$ and $K\Sigma$ photoproduction data. The possible spin and parity of new narrow resonances is discussed.
\end{abstract}


\end{frontmatter}




Compton scattering off the nucleon, as a reaction with subtle convolution of two different scales - electromagnetic and strong interactions, has attracted a lot of attention since a long time. The majority of the investigations found in the literature is devoted to the energy range up to the $\Delta$(1232) region with the aim to explore the nucleon polarizabilities~\cite{Hagelstein:2015egb}. However, the reaction mechanism of Compton scattering beyond the energy region of the $\Delta$(1232) resonance is rarely studied. At high energies, the inelastic channels are emerging through coupled-channel effects and thus essential to the description of Compton scattering, which is dominated by the electromagnetic couplings~\cite{Agashe:2014kda}. About two decades ago, L'vov et al. took advantage of dispersion theory with the help of single-pion photoproduction and resonance photocouplings from an partial wave analysis by which they could extend the range of the model applicability into the second resonance region~\cite{L'vov:1996xd}. More recently, the Giessen coupled-channel model accounted for Compton scattering from the very beginning on, but analyses were limited to the energies below 1.6 GeV due to the lack of the $J = 5/2$ partial waves and resonances at that time~\cite{Feuster:1997pq,Feuster:1998cj,Penner:2002ma,Penner:2002md}. The study in the region of third resonances becomes possible owing to the continuous updates of the Giessen model. The present version accounts for $J = 5/2$ resonances~\cite{Shklyar:2004dy}, careful refinements of isospin $1/2$~\cite{Shklyar:2004ba,Shklyar:2005xg,Shklyar:2006xw,Shklyar:2012js} and $3/2$ partial waves~\cite{Cao:2013psa}, respectively, and includes explicitly $2 \pi N$ channels~\cite{Shklyar:2014kra} based upon the experimental progress from several groups, e.g. CLAS, CBELSA, LEPS, SAPHIR and GRAAL et al.

This objective is reinforced by the very recent measurement of Compton scattering off the nucleon in the center-of-mass (c.m.) energy range of around 1.6 - 1.8 GeV~\cite{Kuznetsov:2010as,Kuznetsov:2015nla}. The observed sharp structures seen in the beam asymmetry data with widths of around 25 MeV is thought to correlate with the narrow enhancement in the data of $\eta$ photoproduction off the nucleon~\cite{Kuznetsov:2006kt,Werthmuller:2015owc} and high-precision measurements of $\pi p$ elastic differential cross sections~\cite{Alekseev:2014pzu,Gridnev:2016dba}. So, Compton scattering off the nucleon in the resonance region is not only a suitable process to study helicity couplings of known resonances, but also an ideal tool to search for possible exotic states that might be weakly coupled to the $\pi N$ state. However, as widely discussed, a solid and comprehensive combined analysis of relevant channels on the ground of available data is highly desirable in order to clarify the underlying nature of those rich spectroscopic structures.



The Giessen model is built on effective Lagrangians, treating mesons and baryons as effective degrees of freedom and obtaining the pion-baryon vertices according to the principles of chiral symmetry. The resulting equations for the scattering amplitudes are solved by a coupled-channel approach respecting gauge invariance. In order to fulfill unitary, the Bethe-Salpeter equation is solved in $K$-matrix approximation,
\bea
T_{fi} = K_{fi} + \textrm{i} \sum_{a,b}K_{fa}\emph{Im}(G_{ab})T_{bi},
\label{eq:Kapp}
\eea
with $i, f$ and $a(b)$ being the initial, final and intermediate states, e.g. $\pi N$, $\gamma N$, $2\pi N$, $\eta N$, $\omega N$, $K \Lambda$ and $K \Sigma$ channels. The kernel $K_{fi}$ includes $s$-, $u$-, $t$-channel and contact terms constructed by effective Lagrangians at the tree-level and the scattering amplitude $T_{fi}$ is easily obtained by solving the reduced multichannel problem. The detailed ingredients of our model Lagrangians and formula of partial-wave decomposition is referred to our previous papers~\cite{Feuster:1997pq,Feuster:1998cj,Penner:2002ma,Penner:2002md}. Here we only outline the strategy for treating Compton scattering of interest.

In our model the Compton scattering kernel $K_{fi}$ is composed of nucleon pole and resonances contributions in the $s$- and $u$-channel, $\pi$- and $\eta$-meson exchange terms in the $t$-channel. In order to retain gauge invariance of the Compton amplitude, the electromagnetic interaction is included perturbatively. This means to treat in Eq.~(\ref{eq:Kapp}) photoproduction channels only to leading order. Hence, the summation over intermediate states $a(b)$ is running only over hadronic states but ignoring the $\gamma N$ state. This approximation makes sense due to the smallness of the electromagnetic couplings constant compared to the hadronic couplings. It has been confirmed that the $\gamma N$ rescattering contributes negligibly little in the $\Delta$-resonance region~\cite{Penner:2002md}. Consequently, the calculation of the hadronic reactions decouples esentially from the electromagnetic ones and can be extracted independently. In practice, Eq.~(\ref{eq:Kapp}) is first solved for the hadronic states only, namely $i, f =$  $\pi N$, $2\pi N$, $\eta N$, $\omega N$, $K \Lambda$ and $K \Sigma$. In an independent second step, the meson photoproduction amplitudes can be extracted by evaluating the T-matrix equation Eq.~(\ref{eq:Kapp}) for the initial channel $i = \gamma N$ but using the previously determined hadronic channels. Finally, the Compton scattering amplitudes are calculated by solving Eq.~(\ref{eq:Kapp}) for  $i, f = \gamma N$. In all steps, the intermediate states $a(b)$ are constrained to purely hadronic channels.

This prescription is non-trivial when the gauge invariance of Compton scattering is considered. The isospin of the photon can be split into an isoscalar part and the third component of an isovector part, respectively. When both initial and final photons are in isovector states, the total isospin could be either $I=1/2$ or $I=3/2$, respectively, thus being weighted differently in the rescattering part of Eq.~(\ref{eq:Kapp}). In the case that gauge invariance for the nucleon contributions is only fulfilled for the proton and neutron amplitude, the gauge invariance of Compton scattering is violated. Alternatively, if we adopt the mentioned perturbative prescription, the Compton isospin amplitudes do not enter in the re-scattering contribution. As a result, gauge invariance of Compton scattering is fulfilled. This also coincides with the fact that only two amplitudes ($\gamma p \to \gamma p$ and $\gamma n \to \gamma n$) are experimentally accessible so only the proton and neutron Compton amplitudes are of interest in the calculation.


Our previous results for the entire set of non-strange and strange channels in $\pi N$ and $\gamma N$ collisions, found in refs.~\cite{Shklyar:2004dy,Shklyar:2004ba,Shklyar:2005xg,Shklyar:2006xw,Shklyar:2012js,Cao:2013psa,Shklyar:2014kra}, are constituting a solid foundation for the present Compton scattering calculations. Not only the data of $\pi N$ collisions are included into these analyses~\cite{Shklyar:2004dy}, but also the most recent data of $\gamma N$ collisions from LEPS, CLAS, MAMI, CBELSA and GRAAL collaborations are used. It is worth to note that the new GRAAL data of proton Compton scattering is not included into the fit in those works. As will be seen in the following, the contribution of isospin $3/2$ sector is dominant in the Compton scattering off the proton in the third resonance region, so the parameters in the isospin $1/2$ sector remains unchanged after including these Compton data as compared to the previous investigation of the $\omega N$~\cite{Shklyar:2004ba}, $K \Lambda$~\cite{Shklyar:2005xg}, and $\eta N$~\cite{Shklyar:2006xw,Shklyar:2012js} final states (see these literatures also for compilation of large amount of data references for these channels). The $\pi^+ p \to K^+ \Sigma^+$, $\pi^- p \to K^+ \Sigma^-$, and $\pi^- p \to K^0 \Sigma^0$ reactions~\cite{Cao:2013psa} are not affected either because here only helicity couplings of relevant resonances are readjusted, as discussed in the following text. The influence on the data of $\gamma p \to K^+ \Sigma^0$ and $\gamma p \to K^0 \Sigma^+$ reactions will be discussed in detail hereafter.

\begin{figure}
  \begin{center}
  {\includegraphics*[width=12cm]{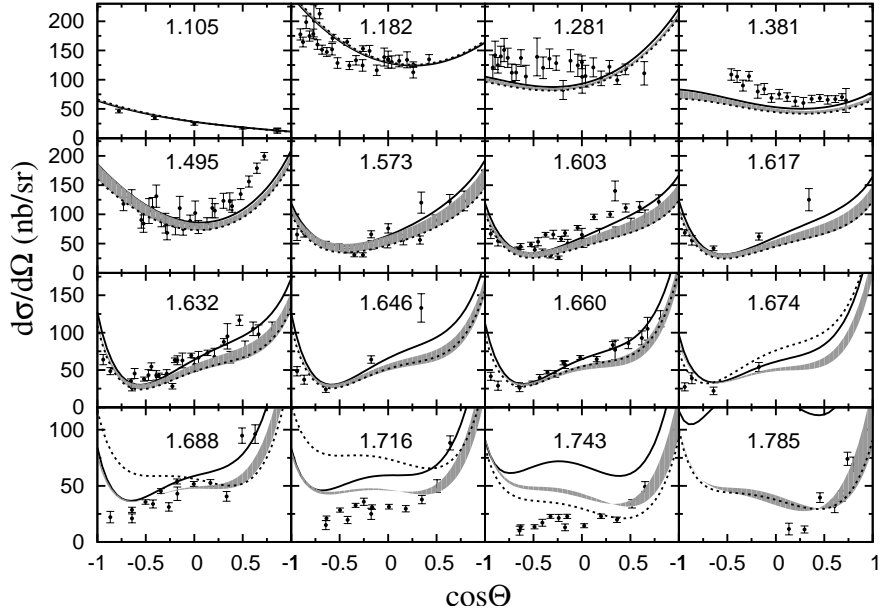}}
       \caption{
  The differential cross sections of proton Compton scattering for different c.m. energies W (in unit of GeV as indicated in each figure). Solid lines: the result with the parameters in Ref.~\cite{Cao:2013psa}; Shaded area: the improved result with adjusting the helicity couplings of $D_{33}$(1700) and $F_{35}$(1905) resonances; Dotted lines: adding $S_{11}$(1680) and $P_{11}$(1720). The data are from Ref.~\cite{data:old,Toshioka:1977wm,Jung:1981wm,Duda:1982uk,Ishii:1979bq,Ishii:1985ei,Wada:1985sh}.
      \label{fig:ggpdif}}
  \end{center}
\end{figure}

\begin{figure}
  \begin{center}
  {\includegraphics*[width=12cm]{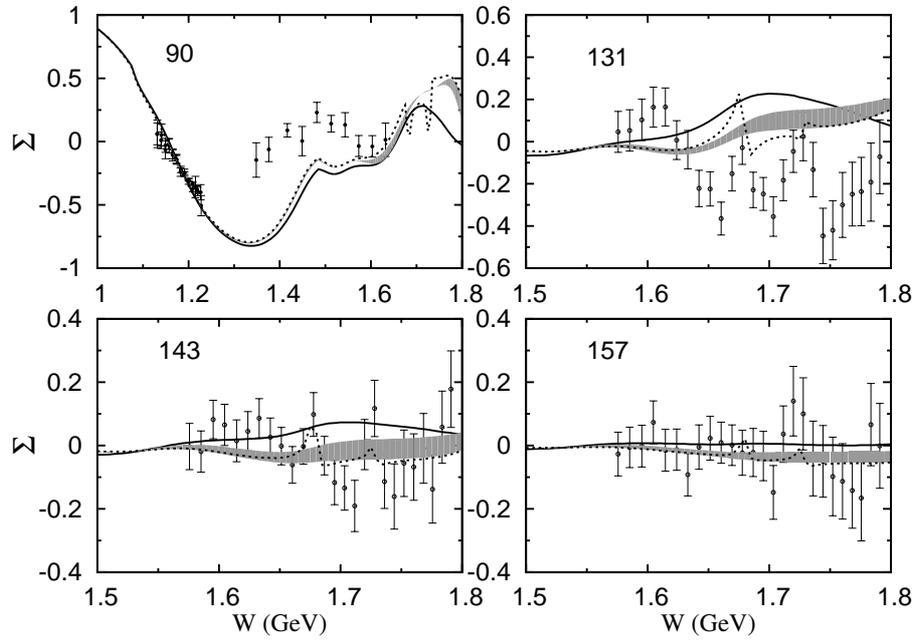}}
       \caption{
  The beam polarization of proton Compton scattering versus c.m. energies W for different angular bins (in unit of degree as indicated in each figure). Line code is the same as in Fig.~\ref{fig:ggpdif}. The data in the $90^{\circ}$ are referred to the compilation in Ref.~\cite{Penner:2002md} and others are from GRAAL~\cite{Kuznetsov:2015nla}.
      \label{fig:ggpsigs}}
  \end{center}
\end{figure}

Our calculated differential cross sections and beam polarization with the parameters in Ref.~\cite{Cao:2013psa} are shown in Fig.~\ref{fig:ggpdif} and Fig.~\ref{fig:ggpsigs}, respectively~\footnote{Those differential cross section data below 1.6 GeV are only selectively illustrated for simplicity. Also, not all available data of beam polarization are shown here, most of which are below 1.2 GeV except those in $120^{\circ}$ angular bin with large error bars. For a full compilation of these old data, please check Ref.~\cite{Penner:2002md} for reference.}. The overall $\chi^2/num.$ is 4.6 with 547 data points in total. The well known $P_{33}$(1232) and $D_{13}$(1520) resonances are dominant below the c.m. energy W$=1.6$~GeV, and our model agrees well with the data in this energy range. The extracted resonance properties are very close to those found in the Particle Data Group (PDG) compilation, as can be seen in Tab.~\ref{tab:para}. This conclusion is also consistent with those of our previous model version and L'vov et al.. As already noted by the latter authors, all other resonance contributions can be safely ignored for energies below 1.6 GeV. The parameters of other resonances with minor contribution can be found in our previous papers~\cite{Shklyar:2012js,Cao:2013psa}. In the $t$-channel, L'vov et al. consider $\pi$- and $\sigma$-meson exchange contributions, but in our model we include $\pi$- and $\eta$-meson exchanges thus avoiding the badly known coupling of the $\sigma$-meson to the di-photon channel. We determined the $\pi\gamma\gamma$ and $\eta\gamma\gamma$ couplings from their well known di-photon decay widths~\cite{Penner:2002md}. These diagrams influence to a minor extent only the backward angle region of the angular distributions.

We obtain a fair description of the data between 1.6 and 1.7 GeV with the caveat that both differential cross sections and beam polarization are fitted well, if we neglect the narrow structures seen in beam polarization data. The main resonance contribution comes from $D_{33}$(1700), whose helicity coupling $A_{\frac{3}{2}}$ is quite close to the central value from PDG, while $A_{\frac{1}{2}}$ is at the lower PDG bound. Its Breit-Wigner ($BW$) mass is much larger than that reported by PDG, but compatible with the partial-wave analysis of the $\gamma p \to \pi^0 \eta p$ reaction of the CB-ELSA Collaboration~\cite{Horn:2008qv}. However, with our parameters in Ref.~\cite{Cao:2013psa} we cannot describe the data in the range of W$=1.7 - 1.8$~GeV, as is obvious from Fig.~\ref{fig:ggpdif}. The obtained $\chi^2/num. = 92.2$ between 1.6 and 1.8 GeV in Tab.~\ref{tab:fit} comes in fact mainly from the range of 1.7 - 1.8 GeV. Besides $D_{33}$(1700), the calculation strongly favor the $F_{35}$(1905) state as another dominating resonance in this third resonance region, whose extracted properties are also listed in Tab.~\ref{tab:para}. The essential role of $D_{33}$(1700) was also discovered by L'vov et al.. However, while we favor $F_{35}$(1905), they stress the importance of the $D_{15}$(1675) and $F_{15}$(1680) resonances instead. The helicity couplings of $D_{15}$(1675) and $F_{15}$(1680) tend to be quite small except the $A^p_{\frac{3}{2}}$ of $F_{15}$(1680), which is a consistent finding in various analyses~\cite{Agashe:2014kda,Shklyar:2006xw,Shklyar:2012js}. The role of $F_{35}$(1905) is determined in our model by a full fit to the recent data of $K\Sigma$ channel and $I = 3/2$ partial waves of $\pi N$ channels so it is more solid and trustful. All the mentioned four resonances are well established independently by different models~\cite{Anisovich:2015tla,Anisovich:2017xqg,Ronchen:2014cna,Kamano:2013iva,Workman:2011vb}, and are assigned a four-star rank by PDG~\cite{Agashe:2014kda}. But in recent beam asymmetry data of $\pi$-meson photoproduction on the proton~\cite{Dugger:2013crn} the helicity couplings of the $D_{33}$(1700) and $F_{35}$(1905) states are found to be very different from the results of the SAID analysis~\cite{Workman:2011vb}.

\begin{table}[t]
  \begin{center}
 \begin{tabular}{|p{1.5cm}<{\centering}|p{0.8cm}<{\centering}|p{1.5cm}<{\centering}|p{0.8cm}<{\centering}|p{1.5cm}<{\centering}|p{1.5cm}<{\centering}|}
 \hline
 $N^*$ or $\Delta^*$ & Ref. & BW mass & $\Gamma_{tot}$ & $A_{\frac{1}{2}}$ or $A^p_{\frac{1}{2}}$  & $A_{\frac{3}{2}}$ or $A^p_{\frac{3}{2}}$ \\
 \hline
   $P_{33}$(1232)
& \cite{Cao:2013psa}& 1227 & 110 & -128 $\pm$ 6 & -253 $\pm$ 8 \\
& PDG               & 1232 & 117 & -135 $\pm$ 6 & -255 $\pm$ 5 \\
 \hline
   $D_{13}$(1520)
&\cite{Shklyar:2012js}& 1505 & 103 &  -15 $\pm$ 1 & 146 $\pm$ 1  \\
& PDG               & 1515 & 115 &  -20 $\pm$ 5 & 140 $\pm$ 10 \\
 \hline
 \hline
   $D_{33}$(1700)
& \cite{Cao:2013psa}& 1673 & 766 & 97          & 147  \\
& this              & 1673 & 766 & 106 $\pm$ 5  & 142 $\pm$ 12 \\
& PDG               & 1700 & 300 & 140 $\pm$ 30 & 140 $\pm$ 30 \\
 \hline
   $F_{35}$(1905)
& \cite{Cao:2013psa}& 1842 & 619 & 54         & -127    \\
& this              & 1842 & 619 & 61 $\pm$ 10 & -78 $\mp$ 15 \\
& PDG               & 1880 & 330 & 22 $\pm$ 5  & -45 $\pm$ 10 \\
 \hline
 \hline
   $S_{11}$(1680)
& this              & 1681 & 2 $\pm$ 1& 32 $\pm$ 10$^\dag$ & --- \\
 \hline
   $P_{11}$(1720)
& this              & 1726 & 2 $\pm$ 1& 35 $\pm$ 10$^\dag$ & --- \\
 \hline
   \end{tabular}
  \end{center}
  \caption{Main parameters used in the paper. The Breit-Wigner (BW) masses and total widthes $\Gamma_{tot}$ are given in MeV. The electromagnetic helicity amplitudes are in the unit of $10^{-3}$ GeV$^{-1/2}$. $^\dag$: sign is not determined.
    \label{tab:para}}
\end{table}

\begin{table}[t]
  \begin{center}
 \begin{tabular}{|p{3.0cm}<{\centering}|p{1.5cm}<{\centering}|p{1.5cm}<{\centering}|p{1.5cm}<{\centering}|p{1.5cm}<{\centering}|}
 \hline
 $\chi^2/num.$                                & data num.  & Solid   & Shaded  & Dotted \\
 \hline
  $\gamma p \to \gamma p$: $d\sigma/d\Omega$  & 95$^*$     & 92.2    & 25.9    & 60.2 \\
 \hline
  $\gamma p \to \gamma p$: $\Sigma$           & 78$^*$     & 1.5     & 2.5     & 1.7  \\
 \hline
  $\gamma p \to K\Sigma$: all                 & 909        & 2.0     & 2.8     & 3.1  \\
 \hline
   \end{tabular}
  \end{center}
  \caption{The obtained fit quality of observables evaluated with $\chi^2$ divided by the number (num.) of data points. The solid, shaded and dotted columns are corresponding to three curves in Fig.~\ref{fig:ggpdif} and Fig.~\ref{fig:ggpsigs}, respectively. $^*$: only data between 1.6 - 1.8 GeV are included.
    \label{tab:fit}}
\end{table}

We find that we can improve the description above W$=1.6$~GeV with reducing the $\chi^2/num.$ in Tab.~\ref{tab:fit} by nearly 2/3 via adjusting a little the helicity couplings of $D_{33}$(1700) and $F_{35}$(1905), which, by the way, also play a significant role in the reactions populating $K\Sigma$ final state. A stringent constraint, however, is defined by requiring that the description of other channels is not spoiled, especially those with $K\Sigma$ final state. This issue is accounted for in the uncertainties in Tab.~\ref{tab:para} with requiring the fit quality of $\gamma p \to K\Sigma$ to vary within 1$\sigma$ band ($\chi^2/num. = 2.8 -2.0 \simeq 0.8$ in Tab.~\ref{tab:fit}). A big portion of the worse of $\chi^2/num.$ is from the polarization observables of the $\gamma p \to K^+\Sigma^0$ reaction. The most obvious change is the considerable increase of the $A_{\frac{3}{2}}$ amplitude for $F_{35}$(1905), which is moved much closer to the PDG value. The $A_{\frac{1}{2}}$ amplitude of $D_{33}$(1700) is also a little raised to the central PDG value. With these parameters, a large improvement in the range of 1.7 - 1.8 GeV is found as depicted by the shaded area in Fig.~\ref{fig:ggpdif} and Fig.~\ref{fig:ggpsigs}. But still, the beam polarization data in the angular bin at $\theta=131^\circ$ is not well described.

As clearly seen in Fig.~\ref{fig:ggpsigs}, the beam polarization data from the GRAAL facility reveal two narrow structures~\cite{Kuznetsov:2015nla}. Those structures are located just above at the position of $K \Lambda$ and $\omega N$ threshold, but we do not obtain any kind of structure at these energies. As a matter of fact, coupled-channel effects generate close to thresholds typically small kinks and cusps rather than such prominent signals. For example, the kink is present near the $\pi N$ threshold, as shown in the $\theta=90^\circ$ angular bin in Fig.~\ref{fig:ggpsigs}. The $K\Lambda$ and $\omega N$ threshold generate kinks at W$=1.61$~GeV and 1.72~GeV in the $\gamma n \to \eta n$ total cross sections~\cite{Shklyar:2012js}.
A global fit to all the data in different channels is performed but a cusp does not appear in the Compton beam polarization.

In the simple Breit-Wigner parametrization of Ref.~\cite{Kuznetsov:2015nla}, the masses and widths of the structures are determined to be around W$=1.681\pm 0.005$~GeV and $1.726\pm 0.005$~GeV and the corresponding widths  are $\Gamma=18\pm 6$~MeV and $\Gamma=21\pm 7$~MeV, respectively. The differential cross sections were measured with acceptable precision at the energies W$=1.688$~GeV and 1.716~GeV a long time ago, which are very close to the nominal peak position of these structures. However, the data do not show any obvious variance when compared to those at nearby energies in Fig.~\ref{fig:ggpdif}, thus missing any hint to sharp resonances. This is in fact an apparent contradiction between the data of differential cross sections and beam polarization. One may expect that it is very unlikely that these contradicting data can be described simultaneously by a theoretical approach. One of probable reasons is that experimental resolution of the incident photon energy is not good enough to observe sharp resonances~\cite{Ishii:1979bq,Ishii:1985ei,Wada:1985sh}. This puzzling situation can only to resolved by further experimental investigation with smaller energy bins. Here we leave this problem behind and pay attention to the GRAAL polarization data.

It is also impossible to determine the spin and parity of these structures with the Compton scattering data at hand. So we look into other reactions for reference. The dip in the differential cross sections at W$=1.68$~GeV in the $\eta$-meson photoproduction is explained by destructive interference between the $S_{11}$(1535) and $S_{11}$(1650) states by our calculations so it is claimed that no strong indication for a narrow state with the mass of around W$=1.68$~GeV is found in the previous analysis~\cite{Shklyar:2012js,Shklyar:2014kra}. This is confirmed by the calculation of chiral quark model~\cite{Zhong:2011ti}. Also the Bonn-Gatchina group finds that the introduction of the narrow resonance $P_{11}$(1680) deteriorates the overall quality of the fit of $\gamma n \to \eta n$ reaction data~\cite{Anisovich:2015tla}. They achieve the same conclusion~\cite{Anisovich:2017xqg} after a full partial wave analysis to the very recent double polarization data from A2 Collaboration~\cite{Witthauer:2016mpi}, which claims a hint for the $P_{11}$ assignment for this structure.
On the other hand, the EPECUR experiment found the evidence of a $S_{11}$ and a $P_{11}$ resonance with the masses (width) of 1688 (17.6) MeV and 1724 (44.2) MeV in the $\pi^- p$ elastic data~\cite{Alekseev:2014pzu,Gridnev:2016dba}, which are very close to the nominal masses and widths of those seen in Compton scattering. For the time being, we assume the first peak is $S_{11}$ and the second one $P_{11}$. With moderate helicity couplings as shown in Tab.~\ref{tab:para}, we can reproduce quite well the structures in Fig.~\ref{fig:ggpsigs}, though as expected the differential cross section deteriorates largely at 1.688 and 1.716 GeV as shown by the $\chi^2/num. = 60.2$ in Tab.~\ref{tab:para}. The significance of new resonances is smaller than 1$\sigma$ significance ($\chi^2/num. =2.5 - 1.7 \simeq 0.8$) as shown in Tab.~\ref{tab:para} due to the big errors of the data. The very small bare Breit-Winger masses ($\sim$ 2 MeV) would indicate that they are indeed strongly affected by nearby particle production thresholds or continua~\cite{Cao:2014vca}. Interestingly, the calculated magnitude of the first peak in Fig.~\ref{fig:ggpsigs} decreases with increasing scattering angle, while that of the second one remains unchanged. This is in accord with the distinguishing feature of polarization data, supporting our choice of the spin and parity of these structures. However, one should be very cautious that this cannot be viewed as evidence for exotic resonances, because these resonances certainly would have noticeable effects in other reactions if we assign their 2 MeV width to any decay channels, which is not confirmed by the data. Other possible mechanisms, e.g. triangle singularity~\cite{Roca:2017bvy}, are waiting for exploration.

In the GRAAL Compton scattering data off quasi-free neutron~\cite{Kuznetsov:2010as} a narrow structure at about W$=1.68$~GeV is present, but the higher one is missing. We refrain from a discussion of the Compton scattering results on the neutron because of the rare photoproduction data off the neutron does not allow a safe determination of the neutron $I=1/2$ resonance helicity couplings $A^n_{\frac{1}{2}}$ and $A^n_{\frac{3}{2}}$~\cite{Shklyar:2006xw}. Moreover, the GRAAL data are uncorrected by the detector efficiency so can not be compared directly to the model calculation.



In summary, the Compton scattering off the proton in the third resonance region is analyzed for the first time within a coupled-channel model. It is the state-of-the-art approach at high energies, to the best of our knowledge. Gauge invariance is perturbatively satisfied and multichannel equations are solved within the K-matrix approximation.  The reaction mechanism is dominant by $P_{33}$(1232) and $D_{13}$(1520) resonances below 1.6 GeV and $D_{33}(1700)$ and $F_{35}(1930)$ above 1.6 GeV. The $\pi N$ and $K \Sigma$ channels are important to settle down the properties of the high-lying isospin $I = 3/2$ resonances. No cusps are generated by the $K\Lambda$ and $\omega N$  threshold effect so the narrow structures in beam polarization from the GRAAL facility can not be explained by this scenario. Conflicting features between these polarization data and the old data of differential cross sections are found. Despite putting aside this conundrum, it is difficult to incorporate the data from all channels, though the description of GRAAL data can be improved with adding two narrow resonances with moderate helicity couplings. Our results are shedding light on the search for missing resonances in the photo-induced reactions off nucleon.

This work is accomplished owing to the previous analysis of the available data of pion- and photon-induced reactions up to c.m. energy of 2.0 GeV simultaneously, where the properties of resonances, especially the  helicity couplings are determined reliably. As a consequence, the available Compton scattering data up to 1.8 GeV are reasonably well described in our model. Our results approve the feasibility of perturbative framework for Compton scattering up to high energies, so it offers a vital way for the future study of Compton scattering off the nucleon in the higher resonance region within various dynamical models once more data will be available.


\section*{Acknowledgments}

We thank Dr. V. Kuznetsov for providing GRAAL data files used in this paper and enlightening discussion on the old data. Useful discussions with Dr. V. Shklyar, Prof. B. S. Zou, and Prof. F. K. Guo in preparing the manuscript is gratefully acknowledged by one of authors (X. C.). This work was supported by the National Natural Science Foundation of China (Grant No 11405222), the Deutsche Forschungsgemeinschaft (CRC16, grant B7 and grant Le439/7) and in part by I3HP SPHERE and the LOEWE.

\end{document}